
\magnification = \magstep0
\nopagenumbers

\hsize=18.6truecm  \hoffset=-1.1truecm  \vsize=23.3truecm  \voffset=-0.1truecm
\hsize=18.6truecm  \hoffset=-0.8truecm  \vsize=24.9truecm  \voffset=-0.7truecm

\input eplain.tex

\font\lc=cmr10
\font\mc=cmr9   \font\mit=cmti9
\font\sc=cmr8   
\font\vsc=cmr7  \font\vsit=cmti7  
\def\mathhexbox#1#2#3{\leavevmode\hbox{$\mathsurround=0pt 
                                       \mathchar"#1#2#3$}}
\def\copyright{{\ooalign
    {\hfil\raise.07ex\hbox{c}\hfil\crcr\mathhexbox20D}}}
\def\xleft{$\phantom{{\rm Vol.}~0}$}
\def\xright{$\phantom{{\rm No.}~0, 199}$}

\def\absbaselines{\baselineskip=11pt \lineskip=0pt \lineskiplimit=0pt}
\def\sglbaselines{\baselineskip=10.4pt \lineskip=0pt \lineskiplimit=0pt}
\def\medbaselines{\baselineskip=10pt \lineskip=0pt \lineskiplimit=0pt}
\def\smlbaselines{\baselineskip=8pt \lineskip=0pt \lineskiplimit=0pt}
\def\vs{\vskip 8pt} \def\vss{\vskip 6pt} \def\vsss{\vskip 2pt}
\parskip = 0pt 
\def\makeheadline{\vbox to 0pt{\vskip-30pt\line{\vbox to8.5pt{}\the
                  \headline}\vss}\nointerlineskip}

\def\footnoterule{\kern-3pt \hrule width \hsize \kern 2.6pt \vskip 3pt}

\def\omit#1{\empty}
\pretolerance=15000  \tolerance=15000
\def\ts{\thinspace}  \def\cl{\centerline}
\def\ni{\noindent}   \def\nhi{\noindent \hangindent=10pt}
       \def\bk{\kern -0.3em}  \def\b{\kern -0.1em}
\def\r0{$\rho_0$}    
\def\0{\phantom{0}}  \def\1{\phantom{1}}  \def\d{\phantom{.}}
\def\etal{{\it et~al.~}} \def\huge{$\phantom{00000000000000000000000000000000}$}
\def\gapprox{$_>\atop{^\sim}$}  
\def\ltapprox{\hbox{$<\mkern-19mu\lower4pt\hbox{$\sim$}$}}
\def\gtapprox{\hbox{$>\mkern-19mu\lower4pt\hbox{$\sim$}$}}

\def\mltapprox{\raise2pt\hbox{$<\mkern-19mu\lower5pt\hbox{$\sim$}$}}

\newdimen\sa  \def\sd{\sa=.1em  \ifmmode $\rlap{.}$''$\kern -\sa$
                                \else \rlap{.}$''$\kern -\sa\fi}
              \def\dgd{\sa=.1em \ifmmode $\rlap{.}$^\circ$\kern -\sa$
                                \else \rlap{.}$^\circ$\kern -\sa\fi}
\newdimen\sb  \def\md{\sa=.06em \ifmmode $\rlap{.}$'$\kern -\sa$
                                \else \rlap{.}$'$\kern -\sa\fi}

\def\kms{km~s$^{-1}$}
\def\s{\ifmmode ^{\prime\prime} \else $^{\prime\prime}$ \fi}
\def\min{\ifmmode ^{\prime} \else $^{\prime}$ \fi}
\def\m31{M{\ts}31}
\def\msun {M$_{\odot}$~}  \def\msund{M$_{\odot}$}  \def\mbh{$M_{\bullet}$}

\parindent=0pt

\headline={\leftskip = -0.15in
           \medbaselines\vbox to 0pt{\sc THE ASTROPHYSICAL JOURNAL, 
           000:L00-L00, 1997 \hfill\null

           \copyright\/ \vsc 1997. The American Astronomical Society. All Rights
           Reserved. Printed in U.S.A. \hfill\null}}

\cl {\null}  \vsss\vsss

\sglbaselines

\cl{SPECTROSCOPIC EVIDENCE FOR A SUPERMASSIVE BLACK HOLE IN NGC 4486B}

\vs

\cl{~J{\sc OHN} K{\sc ORMENDY},$^{1,\ts2,\ts3}$\ts\ts
    R{\sc ALF} B{\sc ENDER},$^2$\ts\ts
    J{\sc OHN} M{\sc AGORRIAN},$^4$\ts\ts
    S{\sc COTT} T{\sc REMAINE},$^{4,\ts5}$\ts\ts
    K{\sc ARL} G{\sc EBHARDT},$^6$}
\cl{D{\sc OUGLAS} R{\sc ICHSTONE,}$^6$
    A{\sc LAN} D{\sc RESSLER},$^7$
    S.~M.~F{\sc ABER},$^8$
    C{\sc ARL} G{\sc RILLMAIR},$^9$
    {\sc AND}
    T{\sc OD} R.~L{\sc AUER}$^{10}$
    }

\vsss
\cl{\it Received 1996 December 17; accepted 1997 March 27}
\vs

\parindent = 37pt
\cl {ABSTRACT}
\vss
{\narrower\absbaselines

\ni\quad The stellar kinematics of the low-luminosity elliptical galaxy NGC 
4486B have been measured in seeing $\sigma_* = 0\sd22$ with the 
Canada-France-Hawaii Telescope (CFHT) and Subarcsecond Imaging Spectrograph.
Lauer and collaborators have shown that NGC 4486B is similar to \m31 in having
a double nucleus.  Here we show that it also resembles \m31 in its kinematics.
Like \m31, NGC 4486B rotates fairly rapidly near the center ($V = 76 \pm 7$ 
\kms~at 0\sd6) but more slowly farther out ($V \simeq 20 \pm 6$ 
\kms~at $r \simeq 4\s$).  Also, the velocity dispersion gradient is very steep:
$\sigma$ increases from $116 \pm 6$ \kms~at $r = 2\s$\ts--\ts\ts6\s to $\sigma
= 281 \pm 11$ \kms~at the center.   This is much higher than expected for an 
elliptical galaxy of absolute magnitude $M_B \simeq -16.8$: even more than \m31,
NGC 4486B is far above the scatter in the Faber-Jackson correlation between 
$\sigma$ and bulge luminosity.  Therefore the King core mass-to-light ratio, 
$M/L_V \simeq 20$, is unusually high compared with normal values for old stellar
populations ($M/L_V = 4 \pm 1$ at $M_B \simeq -17$).

\ni\quad We construct simple dynamical models with isotropic velocity 
dispersions and show that they reproduce black hole (BH) masses derived by more
detailed methods.  We also fit axisymmetric, three-integral models.  Isotropic 
models imply that NGC 4486B contains a central dark object, probably a BH, of 
mass \mbh~$= 6^{+3}_{-2} \times \b10^8$\ts\msund.  However, anisotropic models 
fit the data without a BH if the ratio of radial to azimuthal dispersions is 
$\sim 2$ at $r \simeq$ 1$^{\prime\prime}$.  Therefore this is a less strong BH
detection than the ones in M{\ts}31, M{\ts}32, and NGC 3115.  A dark mass of
$6 \times \b10^8$\ts\msun is $\sim 9$\ts\% of the mass $M_{\rm bulge}$ in stars;
even if \mbh~is somewhat smaller than the isotropic value, \mbh/$M_{\rm bulge}$
is likely to be unusually large.

\ni\quad Double nuclei are a puzzle because the dynamical friction timescales 
for self-gravitating star clusters in close orbit around each other are short.  
Since both \m31~and NGC 4486B contain central dark objects, our results support 
models in which the survival of a double nucleus is connected with the presence
of a BH.  For example, they support the Keplerian eccentric disk model due to
Tremaine.

\vss
\def\huge{$\phantom{0000000000000000}$}

\ni {\it Subject headings:}  {\kern -3pt}black hole physics -- galaxies: 
                             individual (NGC 4486B) --  galaxies: kinematics and
                             \huge dynamics -- galaxies: nuclei

}

\parindent = 10pt

\doublecolumns\sglbaselines

\cl {1.~\sc INTRODUCTION}
\vss

   NGC 4486B is a low-luminosity E1 companion of M{\ts}87.  From the ground, it
looks like a normal dwarf elliptical similar to but slightly brighter than
M{\ts}32.  Faber (1973) suggests that it has been tidally truncated by M{\ts}87;
it may once have been brighter than its present absolute magnitude, $M_B \simeq 
-16.8$ (assumed distance = 16 Mpc).

      NGC 4486B was added to the CFHT BH survey (see Kormendy 1993; Kormendy \&
Richstone 1995, hereafter KR95, for reviews) because {\it Hubble Space 
Telescope} ({\it HST\/}) WFPC1 images showed an elongated, asymmetrical center
(Lauer \etal 1995).  If the double nucleus of \m31 (Lauer \etal 1993) were moved
several times farther away, it would look similarly asymmetric.  Close double 
nuclei are a surprise because their orbital decay timescales are short.  At
least one promising explanation requires a central BH (Tremaine 1995; see 
\S\ts5, below).  This makes NGC 4486B an interesting target for the BH search.


\vs

{\vsc\smlbaselines

$^1${\ts}Visiting Astronomer, Canada--France--Hawaii Telescope, operated by the
    National Research Council of Canada, the Centre National de la Recherche
    Scientifique of France, and the University of Hawaii.

$^2${\ts}Universit\"ats-Sternwarte,{\ts}Scheinerstra\ss e{\ts}1,{\ts}M\"
    unchen{\ts}81679,{\ts}Germany.

$^3${\ts}Institute for Astronomy, University of Hawaii, 2680 Woodlawn Dr., 
    Honolulu, HI 96822.

$^4${\ts}Canadian Institute for Theoretical Astrophysics, University of 
    Toronto, 60 St.~George Street, Toronto M5S 3H8, Canada.

$^5${\ts}Cosmology Program, Canadian Institute for Advanced Research.

$^6${\ts}Dept.~of Astronomy, University of Michigan, Ann Arbor, MI 48109.

$^7${\ts}Carnegie Observatories, 813 Santa Barbara St., Pasadena, CA 91101.

$^8${\ts}UCO/Lick Observatory, Univ.~of California, Santa Cruz, CA 95064.


$^9${\ts}Jet{\ts}Propulsion{\ts}Laboratory, Mail{\ts}Stop{\ts}183-900,
    4800{\ts}Oak{\ts}Grove{\ts}Drive, Pasadena, CA 91109. 

$^{\null{\kern -5pt}10}${\ts}Kitt Peak National Observatory, National Optical
    Astronomy Observatories, P.{\ts}O.~Box 26732, Tucson, AZ 85726.

}

      {\it HST\/} WFPC2 images now prove that NGC 4486B has a double center
(Lauer \etal 1996).  Therefore the result of the present paper -- that it also
contains a massive dark object (MDO) of mass \mbh\ \gapprox \ts$10^8$ \msun -- 
leaves us with two cases of double nuclei found in the presence of MDOs.  This 
suggests that the two phenomena are related.

\vss
\cl {2.~\sc CFHT SIS SPECTROSCOPY}
\vss

     The observations were obtained with the CFHT and the Subarcsecond Imaging 
Spectrograph (SIS).  The slit was aligned with the two nuclei.  Parameters of 
the reduced spectra are given in Table 1.  SIS includes a tip-tilt guider; by
offsetting the guide probe, we centered the galaxy on the slit to 1/4 pixel = 
0\sd04.  An exposure sequence consisted of spectrograph rotation, telescope 
focus, exposures without the grism and slit to center the object, an exposure
with the slit in place but with no grism, the object spectrum, another image 
through the slit but without the grism, one with neither slit nor grism, and a
comparison spectrum.  Between the beginning and the end of a spectral exposure,
the average drift of the galaxy position was 0\sd08 $\pm$ 0\sd01 along the slit
and 0\sd015 $\pm$ 0\sd020 perpendicular to it.

      The seeing was measured on the bracketing direct images.  Brightness
profiles of the galaxy along the slit were measured on these and on the 
spectrum; they agree (Fig.~1), so the PSF was the same for the spectrum and the 
direct images.  Its Gaussian dispersion radius is $\sigma_* = 0\sd28 \pm 0\sd02$
(FWHM $= 0\sd66 \pm 0\sd05$) for the average of four, 1800 s exposures and
$\sigma_* = 0\sd22$ for the best exposure.

      After CCD preprocessing and cosmic-ray cleaning, the Ca II infrared 
triplet region of each spectrum was rewritten on a $\log{\lambda}$ scale.  NGC
4486B is small, so sky spectra from the outer parts of the object spectra were
subtracted.

\vfill\eject

\cl{\null} \vskip 6.63truecm

\includegraphics{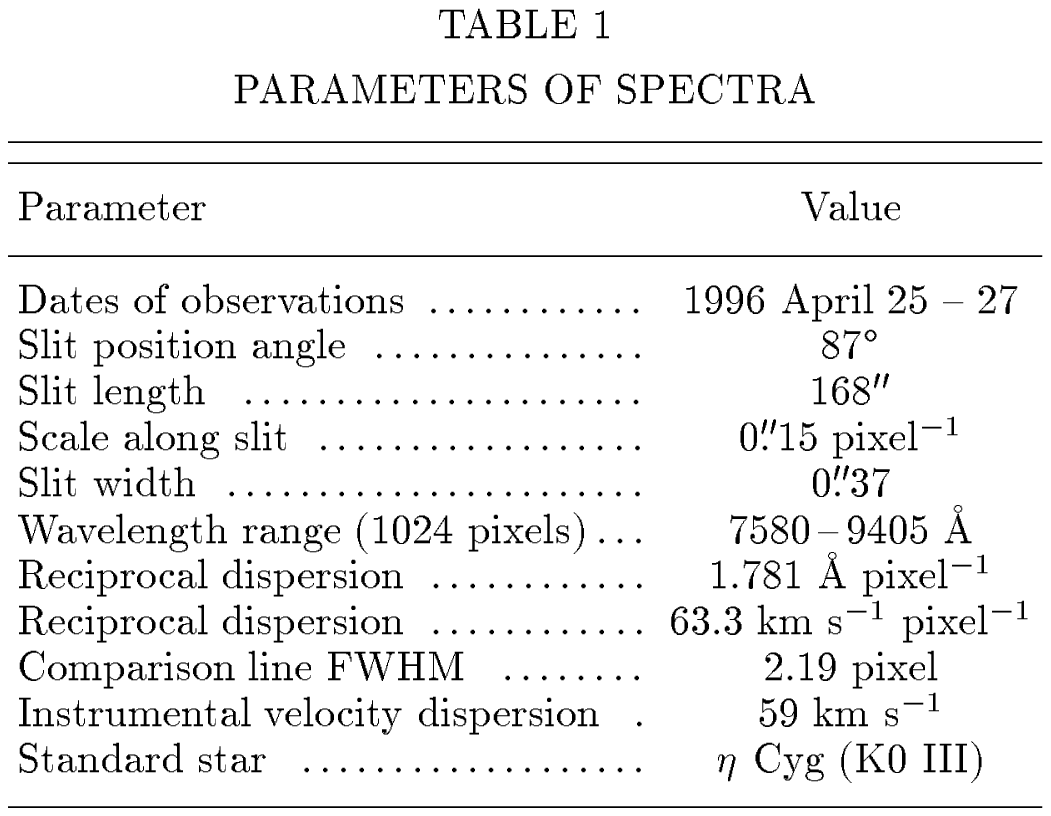}

\headline={\vbox to 0pt{\leftskip = -0.15in
                        \folio\xleft \hfill KORMENDY ET AL. \hfill Vol.~000}}

\vss\vss
\cl {3.~\sc KINEMATIC RESULTS}
\vss

      Velocities and velocity dispersions were calculated with Bender's (1990)
Fourier correlation quotient program.  Results are given in Table 2 and Fig.~1.
The middle section of Table 2 is for the $\sigma_* = 0\sd22$ spectrum; the top
and bottom sections are for the average of all four spectra.

      Since both galaxies have double nuclei, it is interesting to compare NGC
4486B with \m31.  However, note that \m31 is better resolved by the available 
spectroscopy than is NGC 4486B.  In \m31, the separation of the nuclei is 1.8 pc
= 0\sd49 (Lauer \etal 1993), comparable to the best spectroscopic resolution
(FWHM = 0\sd64, Kormendy \& Bender 1996; see KR95) and much larger than the 
spectroscopic scale (0\sd0864 pixel$^{-1}$).  In contrast, the nuclei of NGC
4486B are separated by 12 pc = 0\sd15 (Lauer \etal 1996), i.{\ts}e.~one 
spectroscopic pixel and much less than the spectroscopic seeing, FWHM = 0\sd52
(see Fig.~1).

     NGC 4486B is kinematically similar to \m31.  Both galaxies have rotation
curves that increase toward the center to apparent maximum velocities that are
limited by seeing.  In \m31, $V_{\rm max} = 157 \pm 4$ \kms~and in NGC 4486B,
$V_{\rm max} = 76 \pm 7$ \kms.~\hbox{Outside the spinning center,} $V$ is only 
10\ts--\ts20 \kms~at $r \simeq 4^{\prime\prime}$ in both galaxies.  (At still 
larger radii, \hbox{bulge rotation is important in \m31.)}  Similarly, the 
velocity dispersion at large radii is normal for the luminosity of the bulge 
(see Fig.~2) but then increases rapidly to remarkably large values at the 
center.  Like $V_{\rm max}$, the apparent maximum velocity dispersion is limited
by seeing; it is $246 \pm 8$ \kms~at $\sigma_* = 0\sd27$ in \m31 and $281 \pm 
11$ \kms~at $\sigma_* = 0\sd22$ in NGC 4486B.

      There are also signs of kinematic asymmetries like those in \m31.  The
velocity dispersion is higher on the W side of the center (``P2'' in Lauer \etal
1996) than on the E side.  The velocity zero radius is displaced by 0\sd11 $\pm$
0\sd03 from P2 toward P1.  And the P1 side rotates somewhat more rapidly.
These effects are marginal at the present resolution and should be checked with
{\it HST\/} spectroscopy.

      The central $\sigma$ is much larger than we would expect from the 
Faber-Jackson (1976) correlation between $\sigma$ and $M_B$ (Fig.~2).  The best
stellar-dynamical BH candidates, \m31~(Dressler \& Richstone 1988; Kormendy 
1988) and NGC 3115 (Kormendy \& Richstone 1992; Kormendy {\it et al.} 1996) 
stand out: their central velocity dispersions are above the scatter for other
objects.~NGC 4486B is similarly above the scatter: 
$\sigma$\ts=\ts281\ts$\pm$\ts11 \kms~is surprisingly high for such a 
low-luminosity galaxy. \hbox{Even if NGC 4486B has}
been tidally stripped, $\sigma$ would have been above the scatter
\huge\vskip -10.4pt

\vskip 14.4truecm


\includegraphics{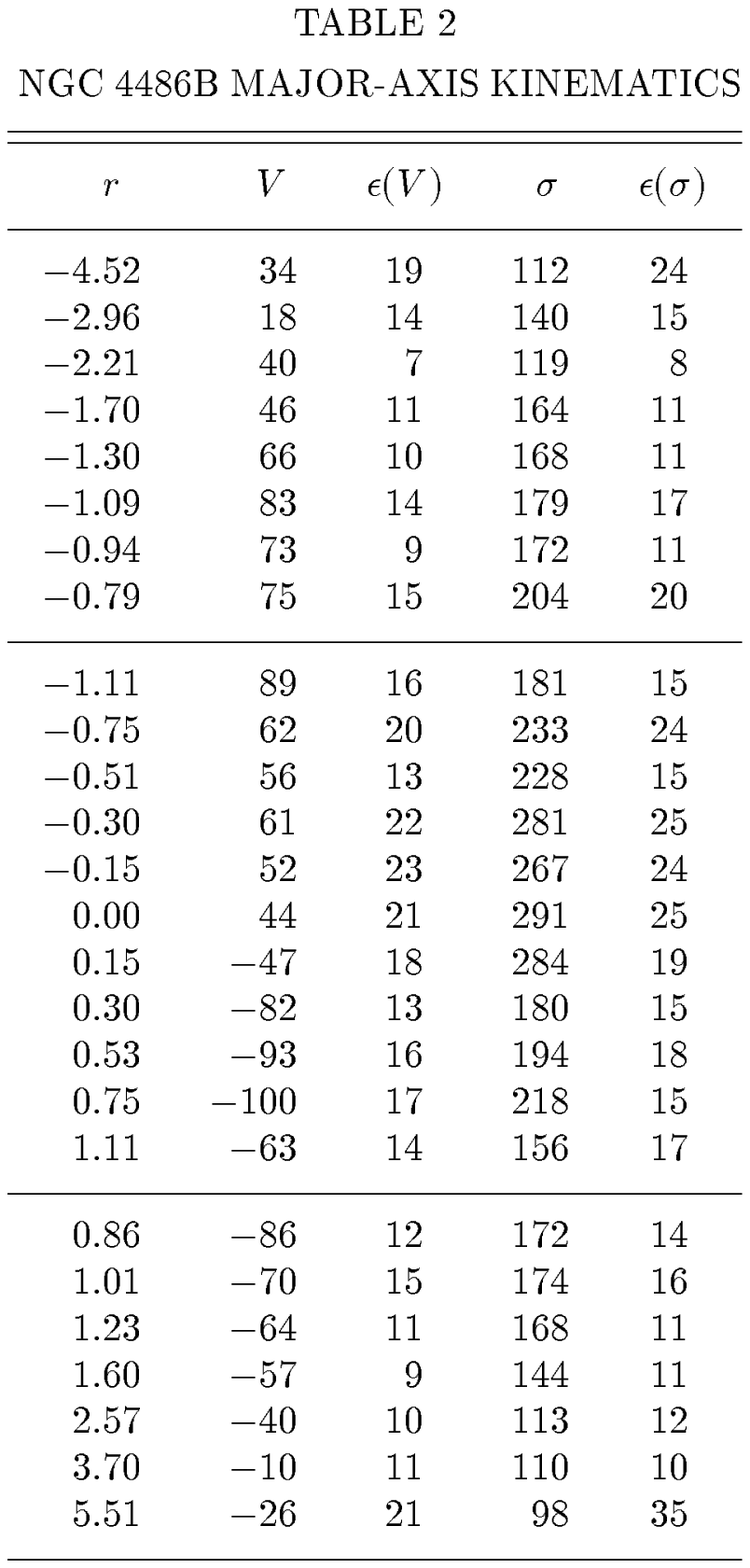}

\ni  of other 
objects for any plausible progenitor luminosity.  Of course, the BH galaxies in
Fig.~2 have been measured with better resolution than most other galaxies.  Our
comments are not meant to imply that these others do not contain BHs; they may 
move upward too when they are observed at high resolution.  But the scatter of
points in Fig.~2 shows the range of $\sigma$ values for which the central 
mass-to-light ratio is normal for an old stellar population.  With respect to 
these, BH candidates stand out as having large $\sigma$.

      For a resolved core, the King core mass-to-light ratio (Richstone \&
Tremaine 1986) provides a good indication of whether there is extra dark matter
near the center.  Where $\Sigma_0$ is the central surface brightness, $r_c$ is 
the core radius, 
and $G$ is the gravitational constant,
$${{M} \over {L}} = {{9 \sigma^2} \over {2 \pi G \Sigma_0 r_c}}.  \eqno (1)$$
Kormendy (1993) emphasizes that $\Sigma_0 r_c \simeq$ constant in different
galaxies, so it is usually not $r_c$ or $\Sigma_0$ that determines whether an
object has a high $M/L$ ratio.  Instead, $M/L$ is high or low if $\sigma^2$ is
high or low compared to other galaxies.  For $\sigma = 281$ \kms, $r_c \simeq 
0\sd23$, and $\Sigma_0$ corresponding to 14.26 $V$ mag arcsec$^{-2}$, the core
mass-to-light ratio is $M/L_V \simeq 20$.  This is much higher than values
($4 \pm 1$ at $M_B \simeq -16.8$) normally observed in old stellar populations.
Therefore NGC 4486B's high velocity dispersion suggests that it contains an MDO.

\vfill\eject

\headline={\vbox to 0pt{\leftskip = -0.15in
            No.~0, 1997 \hfill BLACK HOLE IN NGC 4486B \hfill\xright\folio}}


\cl{\null}\vskip 17.38truecm

\includegraphics{4486b_newfig1.cps}

\vskip -4.8truecm

      {\medbaselines\mc{F{\sc IG}.~1.}---Surface brightnesses and kinematics
along the major axis of NGC 4486B.  In the top panel, the lower solid line is
for the deconvolved {\it HST\/} image (Lauer {\mit et al.}~1996) convolved with
our PSF.  Filled circles show the brightness profile measured from our best 
spectrum.
}


\vskip 7.2truecm

\includegraphics{4486b_fig2.cps}

      {\medbaselines\mc{F{\sc IG}.~2.}---Correlation of $\log{\sigma}$ with
$M_B$ for 594, E and S0 galaxies from McElroy (1995) and for faint ellipticals
from Bender, Burstein, \& Faber (1992).  Distances are based on a Hubble 
constant of $H_0 = 75$ \kms~Mpc$^{-1}$ as in McElroy (1995).  Filled circles 
show apparent central velocity dispersions for \m31~(CFHT: $\sigma_* = 0\sd27$),
NGC 3115 ({\mit HST\/}, aperture = 0\sd21), and NGC 4486B (present data).  Open
symbols are bulge dispersions outside the region affected by the MDO.
}

\cl {4.~\sc SPHERICAL ISOTROPIC DYNAMICAL MODELS}
\vss\vskip 1pt

\def\d{{\rm d}}

      Detailed mass modeling of NGC 4486B is postponed until {\it HST} 
spectroscopy becomes available.  Here we take a first look at whether NGC 4486B
contains an MDO by constructing spherical, isotropic dynamical models.  The 
gravitational potential due to the stars is calculated from the brightness 
distribution, a central MDO is added, and the parameters of both are varied 
until a match is obtained to $(V^2 + \sigma^2)^{1/2}$.  Output are the allowed 
range of stellar mass-to-light ratios $M/L_V$ and MDO masses \mbh.

      The models are constructed as follows.  First, we find the galaxy's
intrinsic light distribution $\nu(r)$ by deprojecting its surface brightness 
profile.  We assume that $M/L_V$ is independent of radius except for \mbh, and 
we derive the model potential $\Phi(r)$ by solving the Poisson equation.  For
each assumed \mbh, we then integrate the Jeans equation,
$$ {\d\over\d r}\left(\nu\sigma^2\right)= -\nu{\d\Phi\over\d r}  \eqno(2) $$
to find the intrinsic second-order moment $\nu\sigma^2(r)$.  We project this and
the zeroth-order moment $\nu(r)$ along each line of sight and convolve the 
results with the spectroscopic seeing.  Finally, we average the seeing-convolved
zeroth- and second-order moments over the bins used in the spectroscopy.
Dividing each second-order moment by the corresponding zeroth-order moment and
taking the square root yields the predicted dispersion profile to be compared 
with the data.  Error estimates on \mbh\ include only the uncertainty in the 
$\sigma$ data; they do not take into account differences between velocity 
measurements and velocity moments, errors in the assumptions (e.{\ts}g., 
anisotropy effects), or distance errors.
\huge

\vskip 8.4truecm

\includegraphics{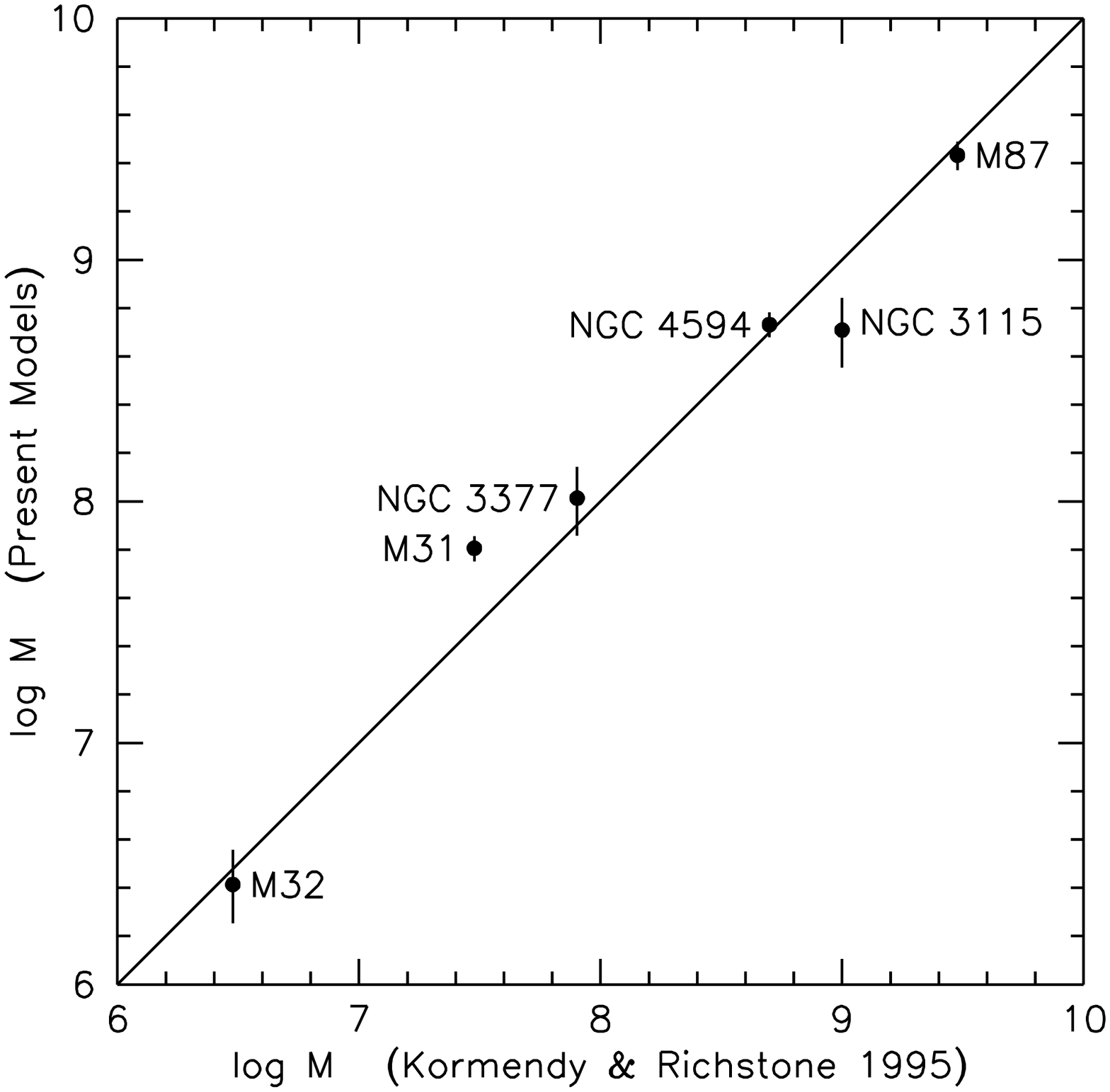}

      {\medbaselines\mc{F{\sc IG}.~3.}---Comparison of BH masses obtained using
the present, simple modeling technique with masses (KR95) derived via more
detailed models (e.{\ts}g., including rotation).} \vs

      Figure 3 compares BH masses derived as above with those obtained from
more detailed models (see KR95 for references).  The highest-resolution 
ground-based results are used; these match the resolution of the present data.
Only M{\ts}32 has been updated since KR95 (Bender \etal 1996).  Figure 3 shows
that for most objects the present technique reproduces the published BH masses
to better than a factor of two.  

\vfill\eject

\vss\vsss\vsss
\cl {5.~\sc A SUPERMASSIVE BH IN NGC \mc4486\sc B}
\vss\vsss\vskip 1pt

      The machinery of \S\ts4 was used to fit the {\it HST\/} WFPC2 photometry
in Lauer \etal (1996) and the kinematic data in Fig.~1.  The results are shown
in Fig.~4.  These models imply that NGC 4486B contains an MDO with \mbh\ $\simeq
8.7 \times 10^8$ \msund.  The corresponding stellar mass-to-light ratio is 
$M/L_V = 5.0$.  As in other MDO detections, the most probable interpretation is 
a supermassive BH. 

\vskip 9.3truecm

\includegraphics{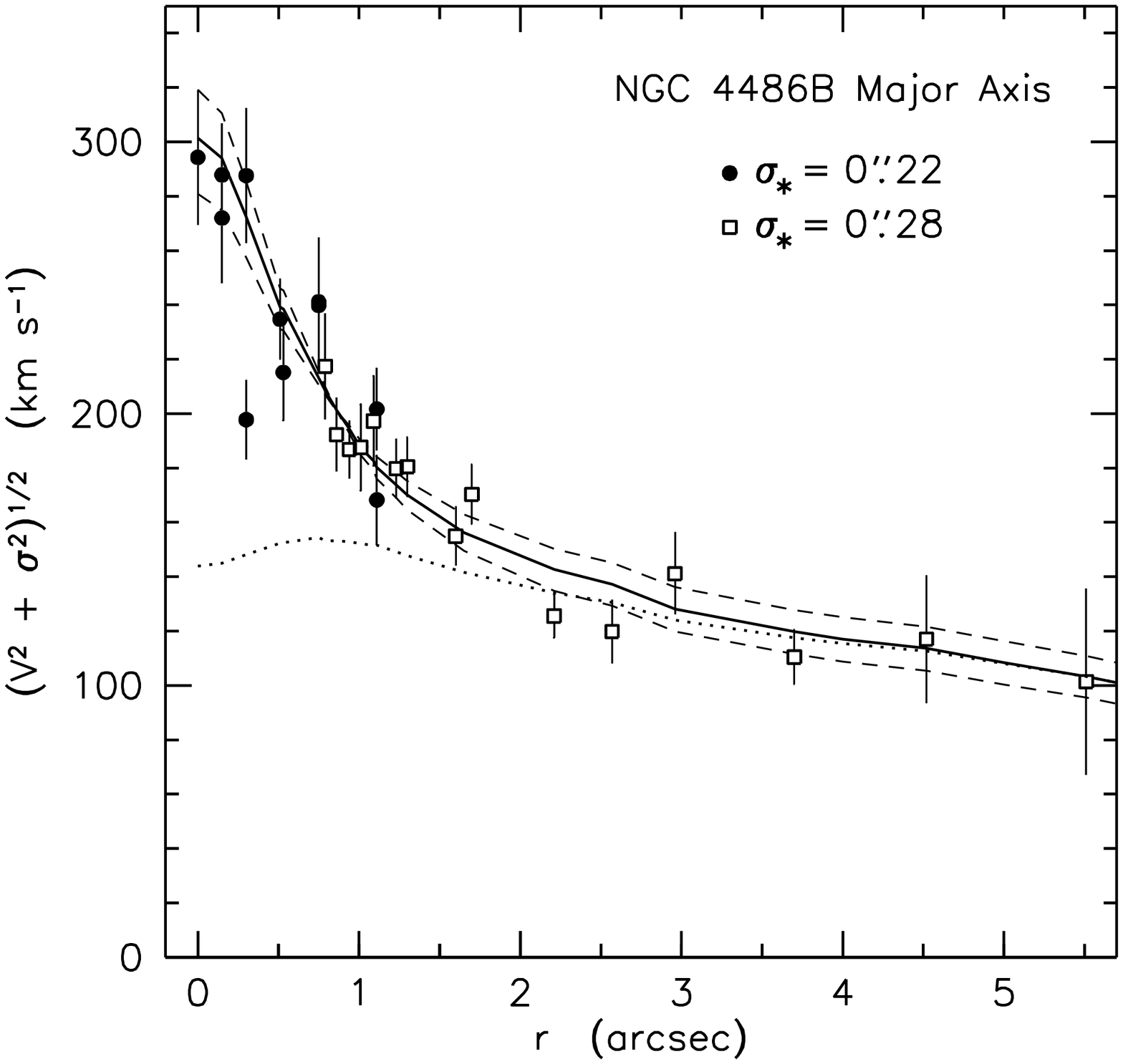}

      {\medbaselines\mc{F{\sc IG}.~4.}---Fits of spherical, isotropic models to
the kinematics of NGC 4486B.  The solid line is the best-fitting model (\mbh\ =
$8.7 \times 10^8$ \msund; $M/L_V = 5.0$).  The dashed lines are ``error bar'' 
models with \mbh\ = $10.5 \times 10^8$ \msund; $M/L_V = 3.8$ and \mbh\ = $6.9 
\times 10^8$ \msund; $M/L_V = 6.3$.  The dotted line fitted at large radii is a
no-BH model with $M/L_V = 6.4$.
}

\vss\vss\vsss

\headline={\vbox to 0pt{\leftskip = -0.15in \lc
                        \folio\xleft \hfill KORMENDY ET AL. \hfill Vol.~00}}

      However, this is a weaker BH detection than the best stellar-dynamical 
cases (the Galaxy, M{\ts}31, M{\ts}32, and NGC 3115).  The reason is that 
axisymmetric, three-integral maximum entropy models (Richstone \etal 1996) can
fit the data without a BH provided that the ratio of radial to azimuthal
velocity dispersions is $\sigma_r/\sigma_\theta \simeq \sigma_r/\sigma_\phi \simeq 2$ at $r \simeq 1^{\prime\prime}$.  Since rotation is relatively
unimportant in NGC 4486B, some anisotropy would not be surprising.  On the other
hand, anisotropic models may be unstable (Merritt 1987).  Also, isotropic model
measurements of \mbh\ have turned out to be close to correct when anisotropic
models were constructed (see KR95 for a review).  E.{\ts}g., in M{\ts}87, the
Sargent \etal (1978) measurement of \mbh\ has turned out to be remarkably
accurate (Harms \etal 1994) even though the galaxy is so luminous that
anisotropy is almost assured and even though anisotropic models can explain the
kinematics without a BH (Duncan \& Wheeler 1980; Binney \& Mamon 1982; Richstone
\& Tremaine 1985; Dressler \& Richstone 1990; van der Marel 1994).  Therefore it
is likely that \mbh\ in NGC 4486B is close to but somewhat smaller than the
value given above.  Even nearly isotropic maximum entropy models give a smaller
mass than those of Fig.~4, i.{\ts}e., \mbh\ $\simeq 4 \times 10^8$ \msund.
In this case, the stellar mass-to-light ratio $M/L_V \simeq 7$ is unusually
high for a galaxy of $M_B = -16.8$, while the models of Fig.~4 imply a more 
nearly normal mass-to-light ratio.  Therefore:

      We adopt the mean of the Jeans and maximum entropy model results, \mbh\
$= 6^{+3}_{-2} \times 10^8$ \msun and $M/L_V = 6 \pm 1$.

      This is a remarkably large BH mass for such a low-luminosity galaxy. It is
well above the correlation between \mbh\ and bulge luminosity that is observed 
for other galaxies (Fig.~5).  For ten previous BH detections, the mean ratio of 
\mbh\ to the mass of the bulge is $<$\null$M_\bullet/M_{\rm bulge}$\null$> = 
0.0022^{+0.0014}_{-0.0009}$.  To date, NGC 3115 has had the largest BH mass 
fraction, 2.4\ts\%.  But $6 \times 10^8$ \msun is $9.4^{+5.0}_{-3.5}$\ts\% of 
the mass in stars.  Even if the galaxy once was brighter, its BH mass is 
surprisingly high.  In fact, NGC 4486B is a good model of what a nearly naked 
quasar (Bahcall \etal 1994, 1995, 1996) would look like after it turned off:
a monstrous BH in a tiny galaxy. However, since anisotropic models allow smaller
BH masses, it will be important to check this result when {\it HST\/}
spectroscopy becomes available.\huge

\huge\vskip -10.4pt
\vbox{\huge
\vskip 7.8 truecm
}

\includegraphics{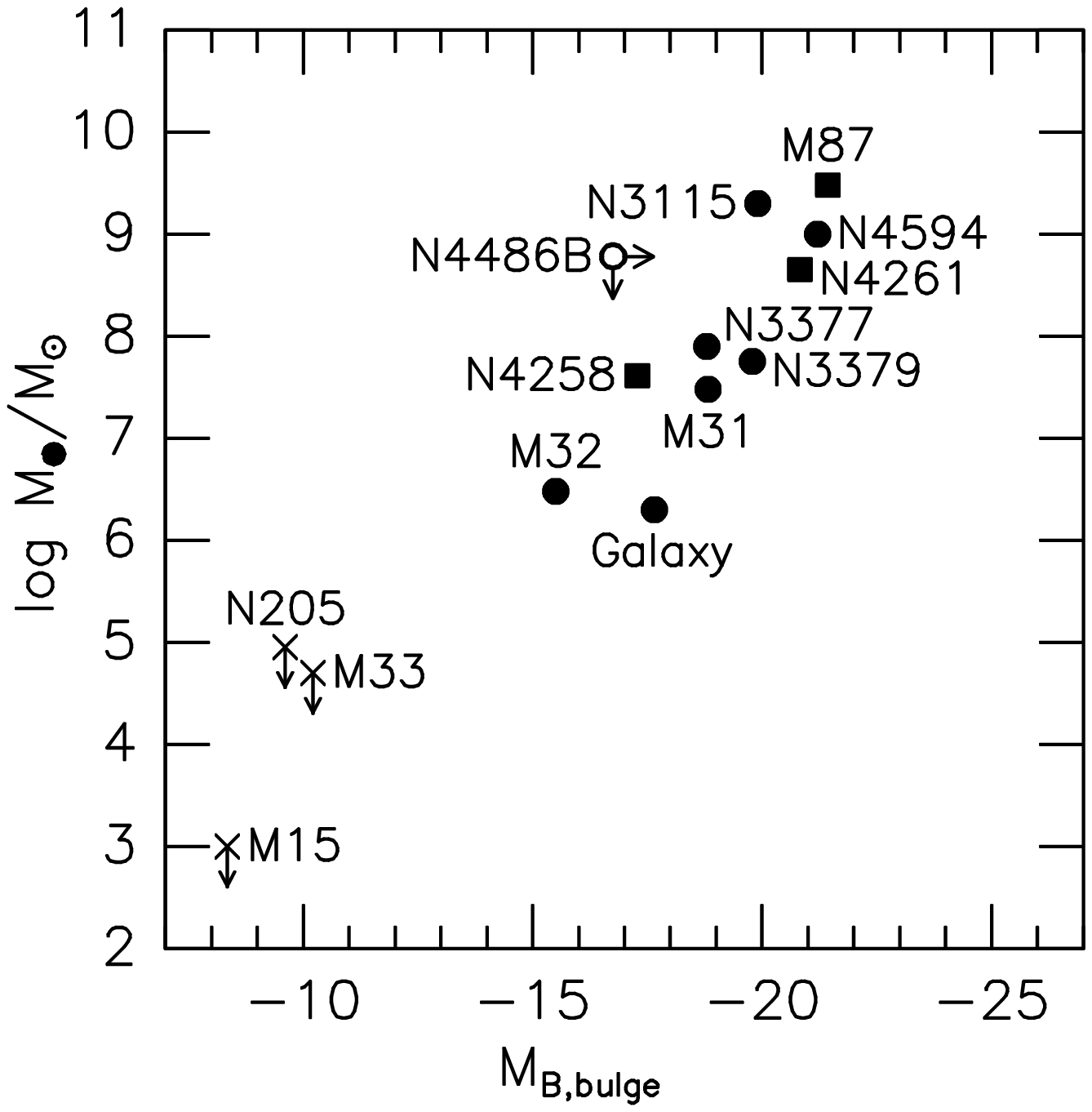}

      {\medbaselines\mc{F{\sc IG}.~5.}---BH mass as a function of bulge absolute
magnitude, from KR95 but with BH masses updated for M{\ts}32, NGC 3115, and NGC
4594, and with NGC 205 (Jones {\mit et al.}~1996), NGC 4261 (Ferrarese {\mit et 
al.}~1996), NGC 3379 (Gebhardt {\mit et al.}~1996) and NGC 4486B (this paper) 
added.   Circles and squares show objects with stellar- and gas-dynamical
evidence for BHs, respectively.  Upper limits on \mbh\ are plotted as crosses.
The correlation may be only the upper envelope of a distribution that extends to
smaller $M_{\bullet}$.
}

\vss\vss\vsss

      An MDO detection in NGC 4486B is interesting because one promising 
explanation for double nuclei requires a BH.  Double nuclei pose an interesting problem because they cannot reasonably be two star clusters in orbit around each
other: the dynamical friction timescale for the decay of their relative orbits
would be short.  Also, for two close double nuclei to appear in the present 
sample of galaxies observed with {\it HST\/}, there should be many more at 
larger separations where merger timescales are still long.  These are not seen.
Tremaine (1995) therefore developed a model of \m31 in which both nuclei are 
parts of the same eccentric disk of stars.  The brighter nucleus (P1), which is
also farther from the barycenter, results from the lingering of stars near 
apocenter, and the fainter nucleus (P2) results from the disk's increase in
density toward the center.  A BH near P2 is required to make the potential 
almost Keplerian.  Only then can an eccentric disk be maintained by modest 
restoring forces such as the disk's self-gravity.

      NGC 4486B now is a second example of a double nucleus.  It, too, contains
an MDO.  This supports Tremaine's model.  The details are different here;
neither nucleus is at the galaxy center.  If the double nuclei are an embedded 
disk, then the disk may have a different orientation or a central hole (Lauer 
\etal 1996).  More generally, our results support models in which the survival 
of a double nucleus is connected with the presence of a BH.

\omit{
      NGC 4486B is also interesting for another reason.  A BH of mass $5 \times
10^8$ \msun is surprisingly large for such a low-luminosity galaxy.  Figure 4 
shows the \mbh\ -- $M_B$ correlation (Kormendy 1993; KR95) with BH masses 
updated and with new galaxies added.  NGC 4261 supports the correlation 
(Ferrarese \etal 1996), which is now based on 9 galaxies.  But the BH mass in
NGC 4486B is a factor of 20 larger per unit luminosity than in the other 
galaxies.
     The remarkable nature of the NGC 4486B BH becomes clear when we look at the
ratio of \mbh\ to the mass $M_{\rm bulge}$ of the bulge.  The latter is obtained
from $M_V$ and from the measured mass-to-light ratio, $M/L_V$.  The mean
BH mass fraction for nine previous detections (those in KR95 plus NGC 4261) is 
$<$\null$M_{\bullet}/M_{\rm bulge}$\null$>$\ts$=$\ts$0.0026^{+0.0017}_{-0.0010}$
(we averaged log\ts$M_{\bullet}/M_{\rm bulge}$).  Omitting the Galaxy, which has
the smallest $M_{\bullet}/M_{\rm bulge}=0.00017$, 
$<$\null$M_{\bullet}/M_{\rm bulge}$\null$>$ $=0.0037^{+ 0.0019}_{- 0.0013}$.
To date, NGC 3115 has had the largest BH mass fraction, $M_{\bullet}/M_{\rm 
bulge}$ = 0.024.  But the remarkable thing about NGC 4486B is that its BH makes
up $\sim 7$\ts\% of the mass of the stars in the galaxy.  Even if it once was 
brighter, its BH mass fraction is surprisingly large.  NGC 4486B clearly is a 
dwarf elliptical with an unusual history.}

\headline={\vbox to 0pt{\leftskip = -0.15in
            No.~0, 1997 \hfill BLACK HOLE IN NGC 4486B \hfill\xright\folio}}

\vss\vss

      We thank D.~McElroy for providing a digital copy of his $\sigma$ catalog.
JK is grateful to the Alexander von Humboldt-Stiftung\ts\ts\ts(Germany) for the 
Research Award that made possible his visit to the Universit\"ats-Sternwarte,
Ludwig-Maximilians-Universit\"at, Munich.  He also thanks the Sternwarte for its
hospitality.  JK's work was supported by NSF grant AST--9219221.  RB's work was
supported by SFB 375 of the German Science Foundation and by the 
Max-Planck-Gesellschaft.  The Nuker team was supported by HST data analysis 
funds through grant GO--02600.01--87A and by NSERC.  We thank the Fields 
Institute for Research in Mathematical Sciences at the University of Toronto for
their hospitality during part of this work.

\huge

\singlecolumn

\vskip -8pt

\vss
\cl{\vsc REFERENCES}
\vss\vskip -5pt

\doublecolumns

{\vsc\smlbaselines

\nhi Bahcall, J.~N., Kirhakos, S., \& Schneider, D.~P.~1994, ApJ, 435, L11

\nhi Bahcall, J.~N., Kirhakos, S., \& Schneider, D.~P.~1995, ApJ, 450, 486

\nhi Bahcall, J.~N., Kirhakos, S., \& Schneider, D.~P.~1996, ApJ, 457, 557

\nhi Bender, R.~1990, A\&A, 229, 441

\nhi Bender, R., Burstein, D., Faber, S.~M.~1992, ApJ, 399, 462

\nhi Bender, R., Kormendy, J., \& Dehnen, W.~1996, ApJ, 464, L123


\nhi Binney, J., \& Mamon, G.~A.~1982, MNRAS, 200, 361






\nhi Dressler, A., \& Richstone, D.~O.~1988, ApJ, 324, 701

\nhi Dressler, A., \& Richstone, D.~O.~1990, ApJ, 348, 120

\nhi Duncan, M.~J., \& Wheeler, J.~C.~1980, ApJ, 237, L27



\nhi Faber, S.~M.~1973, ApJ, 179, 423


\nhi Faber, S.~M., \& Jackson, R.~E.~1976, ApJ, 204, 668

\nhi Ferrarese, L., Ford, H.~C., \& Jaffe, W.~1996, ApJ, 470, 444



\nhi Gebhardt, K., Richstone, D., Kormendy, J., Bender, R., Faber, S., 
      Lauer, T., Magorrian, J., \& Tremaine, S.~1996, BAAS, 28, 1422

\nhi Harms, R. J., {\vsit et al.}~1994, ApJ, 435, L35


\nhi Jones, D.~H., {\vsit et al.}~1996, ApJ, 466, 742


\nhi Kormendy, J.~1988, ApJ, 325, 128 



\huge

\huge

\huge

\huge

\huge

\huge

\huge

\huge

\huge

\huge

\huge

\huge

\huge

\huge

\huge

\huge

\huge

\huge

\huge

\huge

\huge

\huge

\huge

\huge

\huge

\huge

\huge

\huge

\huge

\huge

\huge

\huge

\huge

\huge

\huge

\huge

\huge

\huge

\huge

\huge

\huge

\huge

\huge

\huge

\huge

\huge

\huge

\huge

\huge

\huge

\huge

\nhi Kormendy, J.~1993, in The Nearest Active Galaxies, ed.~J.~Beckman,
     L.~Colina \& H.~Netzer (Madrid: Consejo Superior de Investigaciones 
     Cient\'\i ficas), p.~197



\nhi Kormendy, J., \& Bender, R.~1996, in preparation

\nhi Kormendy, J., {\vsit et al.}~1996, ApJ, 459, L57 

\nhi Kormendy, J., \& Richstone, D.~1992, ApJ, 393, 559

\nhi Kormendy, J., \& Richstone, D.~1995, ARA\&A, 33, 581 (KR95)



\nhi Lauer, T. R., {\vsit et al.}~1993, AJ, 106, 1436

\nhi Lauer, T.~R., {\vsit et al.}~1995, AJ, 110, 2622

\nhi Lauer, T.~R., {\vsit et al.}~1996, ApJ, 471, L79

\nhi McElroy, D.~B.~1995, ApJS, 100, 105

\nhi Merritt, D.~1987, ApJ, 319, 55



\nhi Richstone, D., {\vsit et al.}~1996, in preparation


\nhi Richstone, D.~O., \& Tremaine, S.~1985, ApJ, 296, 370 

\nhi Richstone, D.~O., \& Tremaine, S.~1986, AJ, 92, 72 



\nhi Sargent, W.~L.~W., {\vsit et al.}~1978, ApJ, 221, 731





\nhi Tremaine, S.~1995, AJ, 110, 628





\nhi van der Marel, R.~P.~1994, MNRAS, 270, 271 



}

\vfill\eject

\end